\documentclass[10pt,conference]{IEEEtran}
\IEEEoverridecommandlockouts

\usepackage{cite}
\usepackage{amsmath,amssymb,amsfonts}
\usepackage{algorithmic}
\usepackage{graphicx}
\usepackage{textcomp}
\usepackage{xcolor}
\usepackage{url}
\usepackage{subfig}
\def\BibTeX{{\rm B\kern-.05em{\sc i\kern-.025em b}\kern-.08em
    T\kern-.1667em\lower.7ex\hbox{E}\kern-.125emX}}
   
\def\omit #1{{\textit{OMITTED}}}
\def\omitauthors #1{{\textit{Authors' name omitted for double-blind review}}}
\def\omiturl #1{{\textit{https://omitted-for-review.com}}}
\def\omitacks #1{{\textit{Acknowledgement omitted for double-blind review}}}

\renewcommand{\baselinestretch}{0.97}

\begin{document}

\title{GVSoC: A Highly Configurable, Fast and Accurate Full-Platform Simulator for RISC-V based IoT Processors}
 
\author{\IEEEauthorblockN{Nazareno Bruschi\IEEEauthorrefmark{1}, Germain Haugou\IEEEauthorrefmark{2}\IEEEauthorrefmark{3}, Giuseppe Tagliavini\IEEEauthorrefmark{1}, Francesco Conti\IEEEauthorrefmark{1}, Luca Benini\IEEEauthorrefmark{1}\IEEEauthorrefmark{2}, Davide Rossi\IEEEauthorrefmark{1}}
\IEEEauthorblockA{
\IEEEauthorrefmark{1}\textit{University of Bologna}, Bologna, Italy, \IEEEauthorrefmark{2}\textit{ETH}, Zurich, Switzerland, \IEEEauthorrefmark{3}\textit{GreenWaves-Technologies}, Gronoble, France}
\IEEEauthorblockA{\{nazareno.bruschi, giuseppe.tagliavini, f.conti, davide.rossi\}@unibo.it}
\IEEEauthorblockA{lbenini@iis.ee.ethz.ch, germain.haugou@greenwaves-technologies.com}}
\maketitle

\begin{abstract}
The last few years have seen the emergence of IoT processors: ultra-low power systems-on-chips (SoCs) combining lightweight and flexible micro-controller units (MCUs), often based on open-ISA RISC-V cores, with application-specific accelerators to maximize performance and energy efficiency. Overall, this heterogeneity level requires complex hardware and a full-fledged software stack to orchestrate the execution and exploit platform features. For this reason, enabling agile design space exploration becomes a crucial asset for this new class of low-power SoCs. In this scenario, high-level simulators play an essential role in breaking the speed and design effort bottlenecks of cycle-accurate simulators and FPGA prototypes, respectively, while preserving functional and timing accuracy.
We present GVSoC, a highly configurable and timing-accurate event-driven simulator that combines the efficiency of C++ models with the flexibility of Python configuration scripts. GVSoC is fully open-sourced, with the intent to drive future research in the area of highly parallel and heterogeneous RISC-V based IoT processors, leveraging three foundational features: Python-based modular configuration of the hardware description, easy calibration of platform parameters for accurate performance estimation, and high-speed simulation. Experimental results show that GVSoC enables practical functional and performance analysis and design exploration at the full-platform level (processors, memory, peripherals and IOs) with a speed-up of 2500$\times$ with respect to cycle-accurate simulation with errors typically below 10\% for performance analysis.
\end{abstract}


\section{Introduction}
\label{sec:introduction}
During the last few years, a growing number of research and commercial systems exploit the concept of Parallel Ultra-Low Power (PULP) computing, to tackle performance, flexibility, and energy efficiency requirements of complex near-sensor data analytics applications, including Deep Neural Networks (DNNs) \cite{ojo2018iotdevices}. This computing paradigm aims to exploit heterogeneous system-on-chips (SoCs) composed of state-of-the-art microcontroller unit (MCU) accelerated by tightly coupled clusters of digital signal processors operating near the voltage threshold \cite{pullini2018nearthreshold}.

From a system and application perspective, the exploitation of such extreme heterogeneity requires a full-fledged software stack to orchestrate the execution and exploit platform features. For this reason, enabling fast and accurate design space exploration (DSE), including peripheral subsystems and off-chip components such as memories and sensors, becomes a crucial asset for this new class of low-power systems. In this scenario, high-level simulators play an essential role in breaking the speed and design effort bottlenecks of cycle-accurate simulators and FPGA prototypes, respectively, while preserving functional and timing accuracy. 
The trade-off between the target platform, desired outcomes, application, flexibility, time to simulate, and simulation accuracy determines which simulator use. The most accurate simulators are typically slow because they simulate cycle-by-cycle using the entire platform description \cite{yi2006simulationofcomputer}, often written in Hardware Description Language (HDL). In contrast, the fastest simulators typically simulate only the functionalities of a system but without timing and performance information \cite{chiou2007fast}. In the middle, several simulators provide reasonable fast simulation and accuracy but typically without simulating the entire system and the possibility to quickly explore a range of different configurations, crucial for DSE \cite{elrabaa2017chipmultiprocessor}.

In this work, we present GVSoC, an open-source simulator\footnote{\url{https://github.com/pulp-platform/pulp-sdk}} targeting the PULP architectures that can simulate complex full-platforms, including multicore, multi-memory levels (i.e., on- and off-chip), multi I/O peripherals, and accelerators, laying the bases of a real DSE for IoT embedded systems, with $<$10\% of error with respect to physical embodiments of the target architecture. Besides this high accuracy, GVSoC is up to 2500$\times$ faster than strictly cycle-accurate simulators, guaranteeing the fastest and reliable way to explore new architecture features.
To demonstrate the performance, accuracy, and flexibility of the proposed simulator, we propose several examples of real-life applications that exploit the key features of this simulator.

In Section~\ref{sec:related_work}, we analyze the state-of-the-art simulators the embedded systems, providing a comprehensive overview of the main techniques to simulate platforms.
Then, in Section~\ref{sec:target_arch} we discuss the PULP architecture.
Section~\ref{sec:gvsoc_architecture} details the GVSoC features, describing the internal architecture and the tools that allow the users to inspect the execution.
In Section~\ref{sec:use_cases}, we illustrate how GVSoC can be used to simulate real applications, evaluating its accuracy and simulation performance with respect to an hardware-accurate FPGA-based full-SoC emulator. Finally, in Section \ref{sec:state_of_the_art} we compare the results and the features of GVSoC with other widely spread simulators.

%
%%%%%%%%%%%%%%%%%%%%%%%%%%%%%%%% RELATED WORKS %%%%%%%%%%%%%%%%%%%%%%%%%%%%%%%%%%%%%%%%%
%
\section{Related work}
\label{sec:related_work}
Simulators can be classified based on \textit{i}) their level of abstraction in modeling the target platform; and \textit{ii}) their specialization for target systems and metrics such as multiprocessor/multicore \cite{fu2014prime}, energy/power \cite{walker2018energy}, and accelerator simulators \cite{ankit2019puma}.
Along the first axis, we can distinguish two major classes: functional and timing simulators \cite{akram2019survey}: in this section, we survey the main simulation approaches with a focus on simulators for RISC-V systems.

\subsection{Functional Simulators}
\label{sec:related_work/functional_simulators}
Functional simulators aim at replicating functional behavior, without modeling the internal details of the target architecture. For instance, a functional simulator can model just the Instruction Set Architecture (ISA) of the core and not its pipeline. For this reason, they are typically very fast but do not model the different micro-architectural parameters and do not support accurate performance assessment.
A notable example of simulators in this class is Spike\footnote{\url{https://github.com/riscv/riscv-isa-sim}}, which is the golden reference functional software simulator for RISC-V architecture. It is written in C++ and can be used as a starting point for running software on a RISC-V target. SimpleCPU of gem5 \cite{binkert2011gem5} includes a model for atomic memory accesses, which, after a memory request, returns an approximate time to complete the request without simulating contentions and queuing. This information is used to estimate overall cache access time \cite{simplecpu}.
Binary translators follow different approach, converting, statically or dynamically, one or more basic blocks from the simulated ISA to the host native code, emulating the target architecture. QEMU is a widely used binary translator that can also be modified to collect meta-data from the execution, such as memory accesses \cite{bellard2005qemu}. RISCV-QEMU\footnote{\url{https://github.com/riscvarchive/riscv-qemu}} is the version of QEMU that supports RISC-V target architecture.

Although functional simulators can be used to develop runtime frameworks, compilers, and firmware, their poor accuracy prevents their adoption for DSE and application optimizations.

\subsection{Timing Simulators}
\label{sec:related_work/timing_simulators}
Unlike functional simulators, timing simulators model the micro-architecture of the target. Thanks to their lower level of system abstraction, the modeling accuracy of the target architecture increases, including cache hierarchies, core pipelines, predictors, DMA, and more other micro-architectural features. Moreover, also this category requires a further distinction among cycle-accurate, instruction-level and event-driven \cite{akram2019survey}.
\paragraph{Cycle-accurate}
The simulation proceeds with an accurate cycle-by-cycle approach, using a hardware description language (HDL) such as System Verilog, translated with tools such as closed-source ModelSim or open-source Verilator. ModelSim is a source-level verification tool \cite{modelsim}, while Verilator converts System Verilog description to C++, increasing parallelism and multi-threading and executing it on host hardware like an executable file \cite{verilator}. Overall, these simulators are as hard to extend or implement new features as directly in the hardware and have low simulation speed than the other kind of simulators, running in the kMIPS range \cite{akram2019survey}.
%
%%%%%% PULP %%%%%%
\begin{figure*}[t!]
  \centering
\includegraphics[scale=0.3]{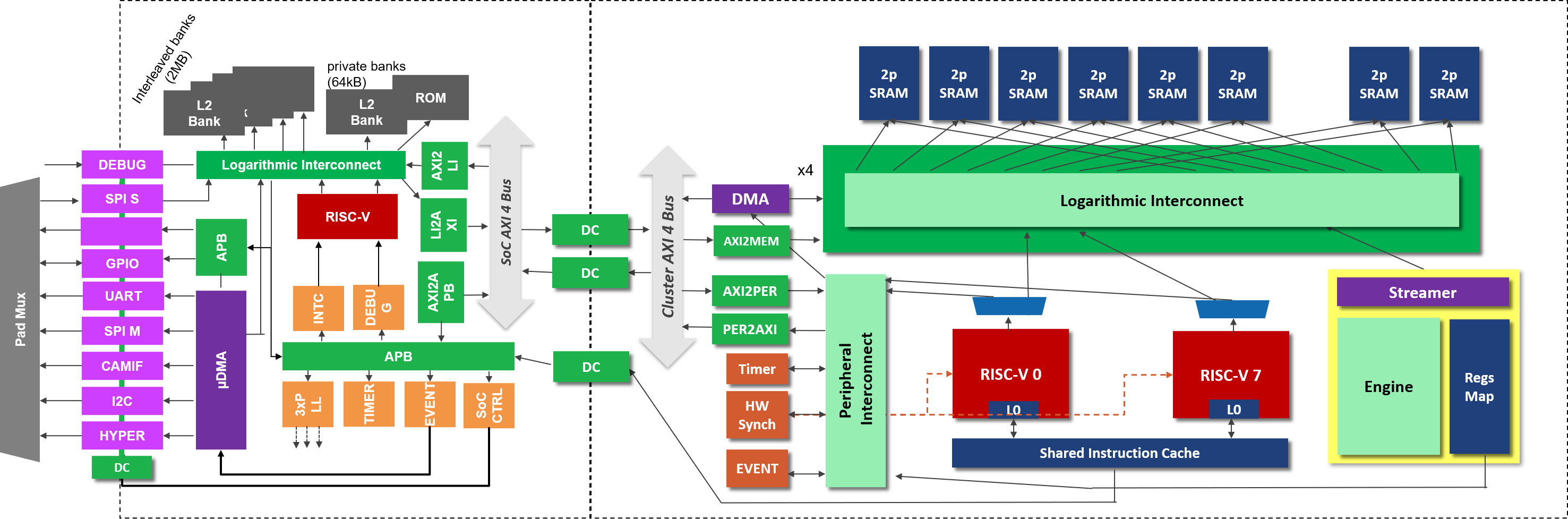}
  \caption{Target architecture. On the left the PULP SoC and on the on the right the PULP CL}
\label{fig:target_arch/pulp_arch}
\vspace{-0.25cm}
\end{figure*}
%%%%%% PULP %%%%%%
%
\paragraph{Instruction-level}
The simulation is conducted imitating the target system, instruction-by-instruction, but modeling the hardware with a coarse-grain level of abstraction (i.e., not in a cycle-accurate way). Instruction-level simulators are faster than cycle-accurate but still relatively slow with respect to the event-driven ones \cite{akram2019survey}. The O3 CPU model of gem5 adopts a model that executes the instructions in the execute pipeline stage \cite{o3gem5}. Another example in gem5 is the TimingSimple model \cite{simplecpu}, which gives up micro-architectural details to boost performance, assuming one cycle for every non-memory instruction and modeling accurately only memory access instructions.
\paragraph{Event-driven}
These simulators do not directly consider cycles but the more abstract concept of \emph{event}, i.e., a change of state in the system occurring at a specific instant in time.
Hardware models schedule events in a queue based on their latency, and no change in the system is assumed to occur between consecutive events.
With this approach, the simulator can directly jump to the occurrence time of the next event in the queue, with a consistent saving in simulation time.
SystemC is a Domain-Specific Language (DSL) built upon C++ classes and helper macros that provides an event-driven simulation environment.
In addition to its basic features for system-level modeling, SystemC provides a Transaction Level Modeling (TLM) layer to model the interface between interoperable modules \cite{systemc}.
RISC-V-TLM \cite{monton2020riscv} is a RISC-V simulator based on SystemC+TLM, including an ISS, a memory, and a basic set of peripherals.
GVSoC adopts a similar approach, and its accuracy is comparable with a SystemC+TLM model.

Timing simulators are often enough to explore parts of the system. However, they require a long time to simulate a real-life end-to-end application in state-of-the-art optimized platforms or have insufficient accuracy. GVSoC, like other event driven simulators, provides an advantageous trade-off between simulation speed, timing accuracy, and completeness; while at the same time enabling highly flexible simulation of the full-systems, justifying its use for DSE with realistic workload and I/O interactions.
%
%%%%%%%%%%%%%%%%%%%%%%%%% TARGET ARCHITECTURE %%%%%%%%%%%%%%%%%%%%%%%%%%%%%%%%%%
%
\section{Target Architecture}
\label{sec:target_arch}
The proposed simulator targets systems implementing the architectural template defined by the Parallel Ultra-Low Power (PULP) platform \cite{pulp}. PULP is an open-source\footnote{\url{https://github.com/pulp-platform}} heterogeneous computing platform consisting of a state-of-the-art micro-controller unit (PULP SoC) accelerated by a parallel programmable accelerator, as shown in Fig.\ref{fig:target_arch/pulp_arch}.

The PULP SoC is hosted by a RISC-V processor called Fabric Controller (FC), managing the peripherals subsystem and offloading compute-intensive tasks to a parallel computing cluster (PULP CL). The PULP SoC is equipped with a main memory typically ranging from 256kB and 2 MB of SRAM (L2), hosting the resident code and application data. The PULP SoC features a complete set of micro-controller peripherals, including JTAG, SPI, I2C, I2S, GPIOs, as well as an HyperBUS DDR interface used to connect to external IoT DRAM or Flash memories such as Cypress Semiconductor HyperRAM/FLASH \cite{flash/ram} or APMemory IoT RAMs \cite{apmemory}. An I/O DMA (micro-DMA), integrated inside the I/O subsystem, performs autonomous data transfers between the L2 memory and the peripherals, once programmed by the FC. Once the transfer is finished, the micro-DMA notifies the end of the transfer to an interrupt controller, which notifies the core accordingly to the pre-configured policy. Typically, in PULP platforms, the PULP SoC, the peripherals, and the PULP CL reside in three different clock domains so that the frequency of each domain can be tuned to sustain the application workload with low power consumption (typically up to 500 MHz for a 22nm technology node). The clocks of the peripherals can be further divided in order to match the operating frequency of slower external devices.

PULP CL consists of a parametric number of identical RISC-V cores sharing a multi-banked Tightly Coupled Data Memory (TCDM), whose size and the number of banks can be configured as well (typically between 64 and 256kB). The TCDM banks can be accessed in a single clock cycle through a low-latency logarithmic interconnect, enabling fast data sharing among processors for data-parallel applications. TCDM, like L2, addresses the banks in a word-level interleaved fashion (i.e., 4B per bank), reducing the probability of contentions between simultaneous core accesses. The cores fetch instructions from a 2-level instruction cache composed of a small private L1 cache (e.g., 512B - 1kB) refilling from a larger shared L1.5 cache (e.g., 4-8kB) within the cluster. The L1.5 cache refills from the L2 memory through a 64-bit AXI bus. A multi-channel DMA autonomously moves the data from L1 to L2 and vice versa. At the same time, a hardware synchronization unit implements in hardware common hardware synchronization functions such as barriers, thread dispatching, and critical sections. The cores within the PULP CL support DSP extensions targeting energy efficient digital signal processing. These ISA extensions, called Xpulp, have several versions \cite{gautschi2017riscy}\cite{garofalo2020xpulpnn} and are updated across SoC generations to better match the target application domain. Floating-point units can be shared among the cores, where the degree of sharing is an architecture parameter, depending on the area and power constraints. A standard interface, named HWPE\footnote{https://hwpe-doc.readthedocs.io/en/latest/}~\cite{conti2013}, allows to include specific hardware accelerators in the PULP CL domain, sharing L1 memory (i.e., the TCDM) with general-purpose cores.
%
%%%%%%%%%%%%%%%%%%%%%%%%% GVSOC ARCHITECTURE %%%%%%%%%%%%%%%%%%%%%%%%%%%%%
%
\section{GVSoC Architecture}
\label{sec:gvsoc_architecture}
%
%%%%%%%% STRUCTURE %%%%%%%%
\begin{figure*}[t!]
\centering
\subfloat[\label{fig:gvsoc_architecture/gvsoc_structure}]{\includegraphics[scale=0.34]{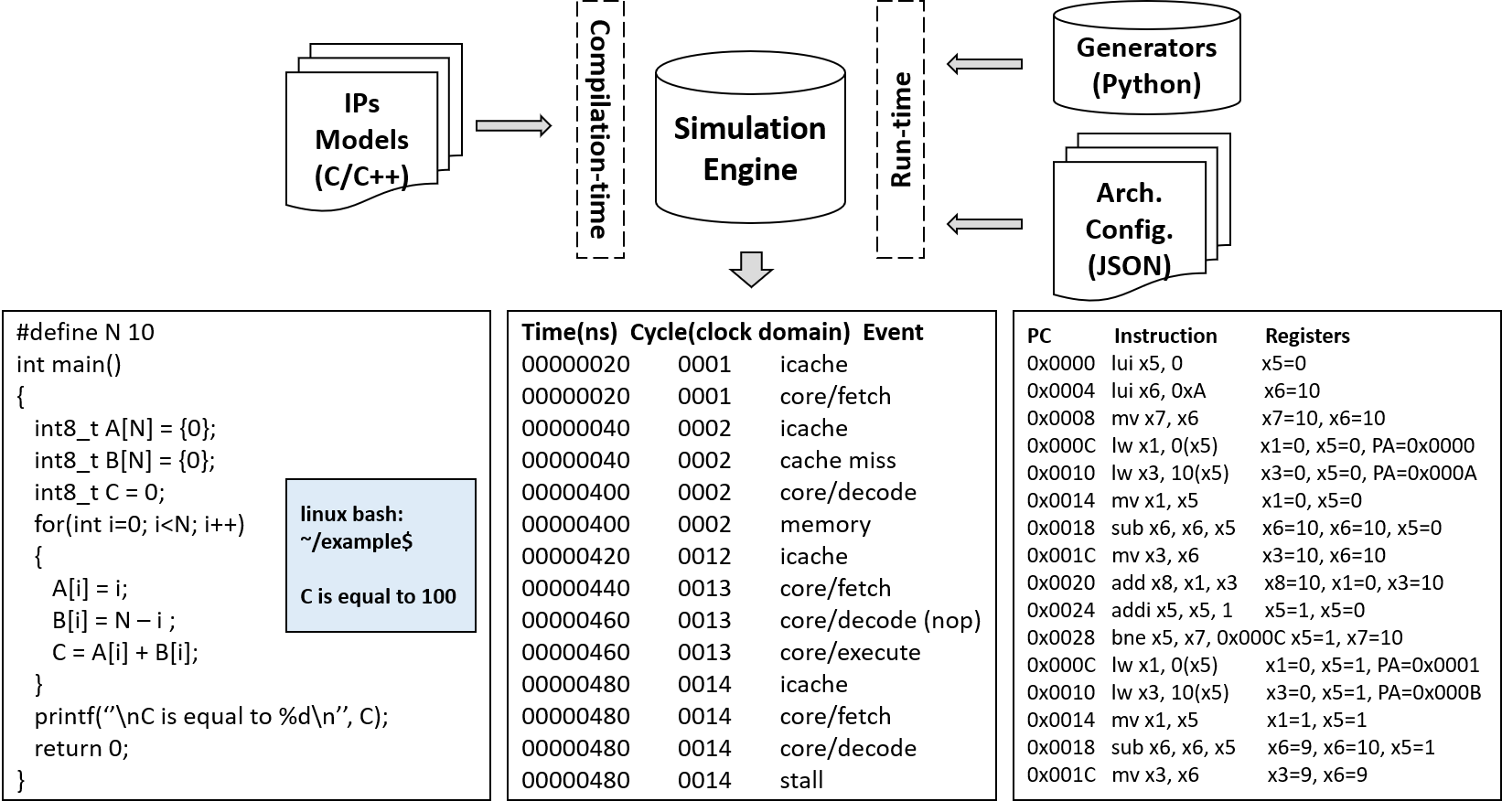}}
\hfill
\subfloat[\label{fig:gvsoc_architecture/gvsoc_req}]{\includegraphics[scale=0.28]{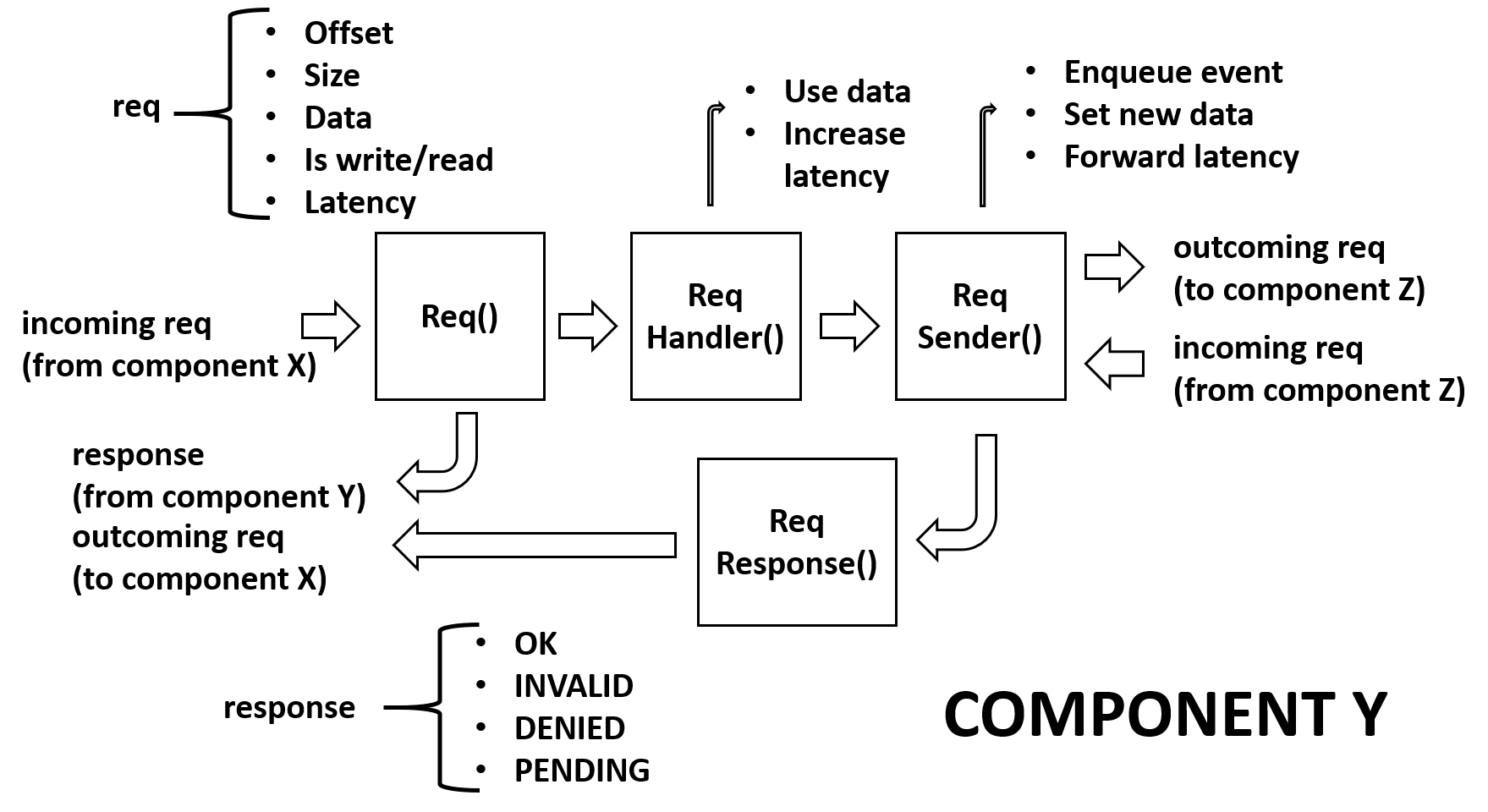}}
\caption{GVSoC structure and features. (a) Main parts that compose GVSoC: JSON files describe the architecture, Python generators instantiate the components and C++ models describe all the IPs (b) How GVSoC components interact with each other. Every component receives a request containing the information to forward it to another component or to handle it by itself.}
\label{fig:gvsoc_architecture/gvsoc_internally}
\vspace{-0.3cm}
\end{figure*}
%%%%%%%% STRUCTURE %%%%%%%%
%
%%%%%%%% BUFFERS %%%%%%%%
\begin{figure}[t!]
  \centering
\includegraphics[scale=0.37]{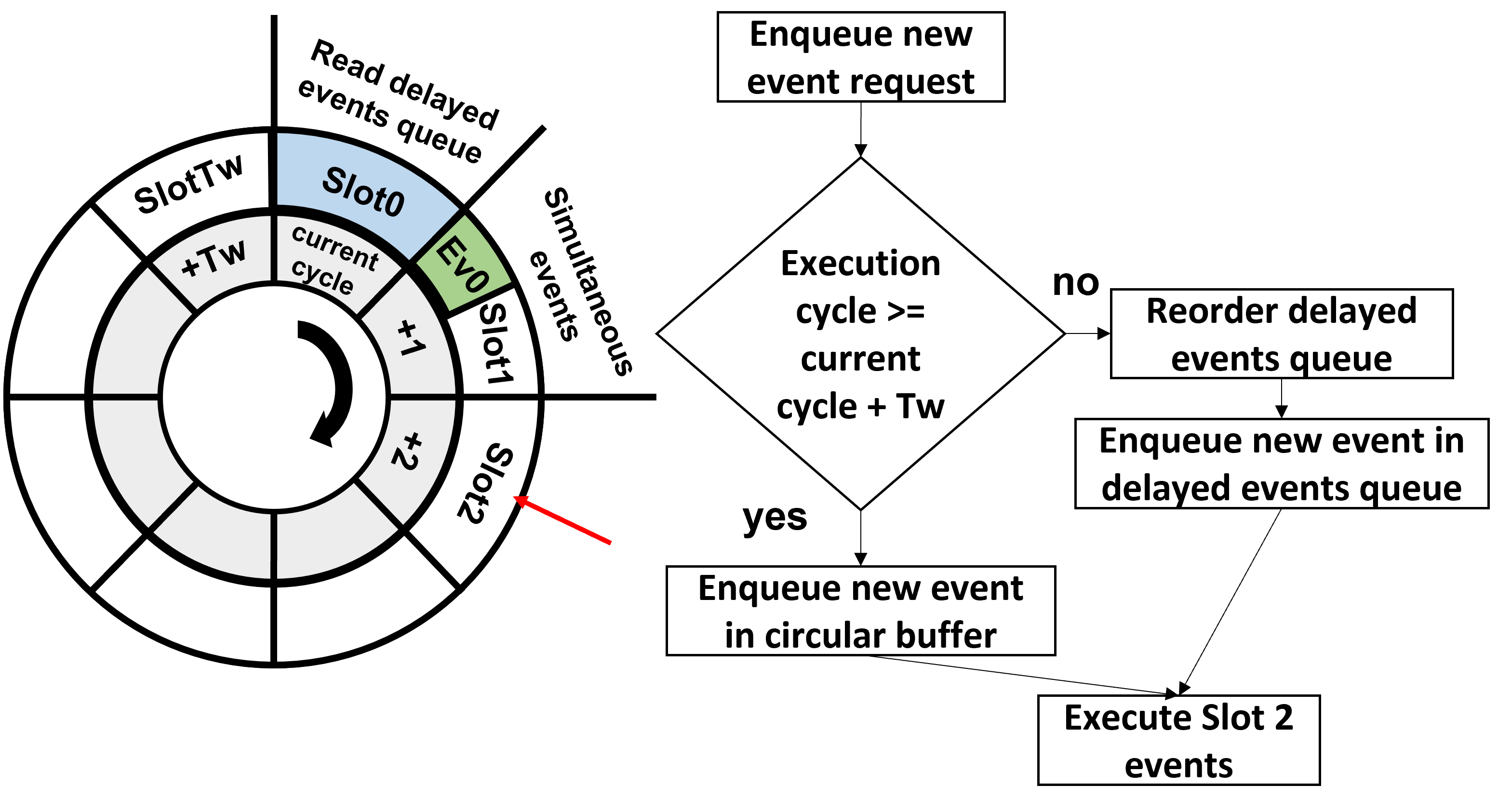}
  \caption{GVSoC events management. A circular buffer contains every enqueued event to be executed in a fixed time window Tw, and its positioning algorithm}
\label{fig:gvsoc_architecture/time_modelling}
\vspace{-0.25cm}
\end{figure}
%%%%%%%% BUFFERS %%%%%%%%
%
%%%%%%%% SYNCH %%%%%%%%
\begin{figure*}[t!]
\centering
\subfloat[\label{fig:gvsoc_architecture/gvsoc_clock_domain_cross}]{\includegraphics[scale=0.4]{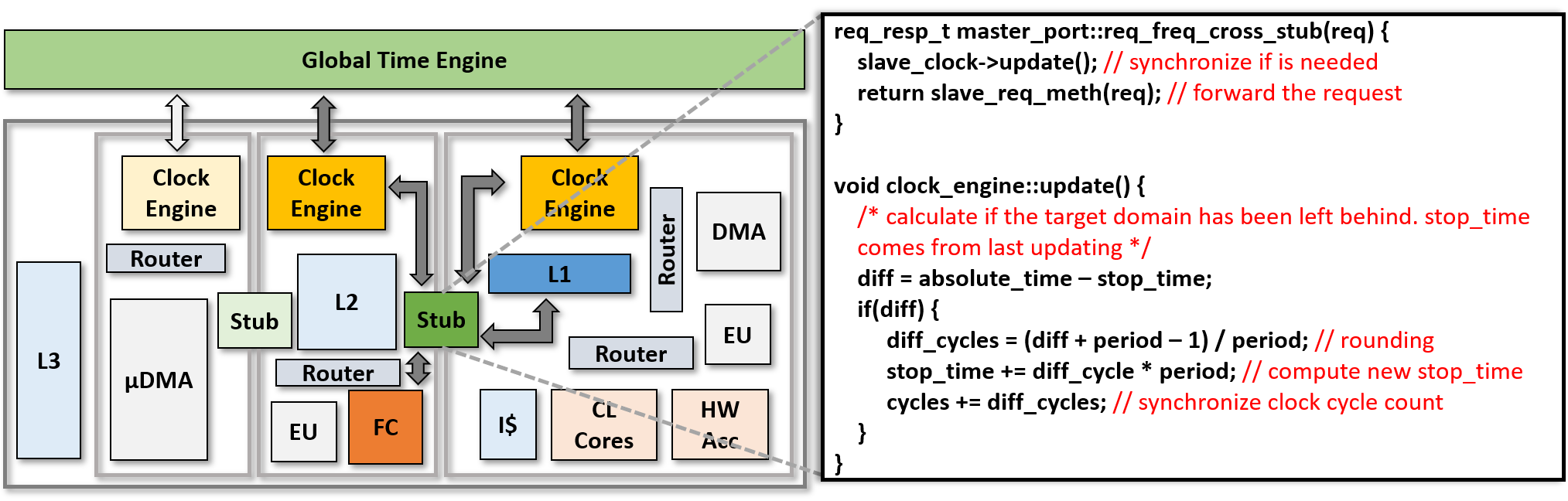}}
\hfill
\subfloat[\label{fig:gvsoc_architecture/counters}]{\includegraphics[scale=0.26]{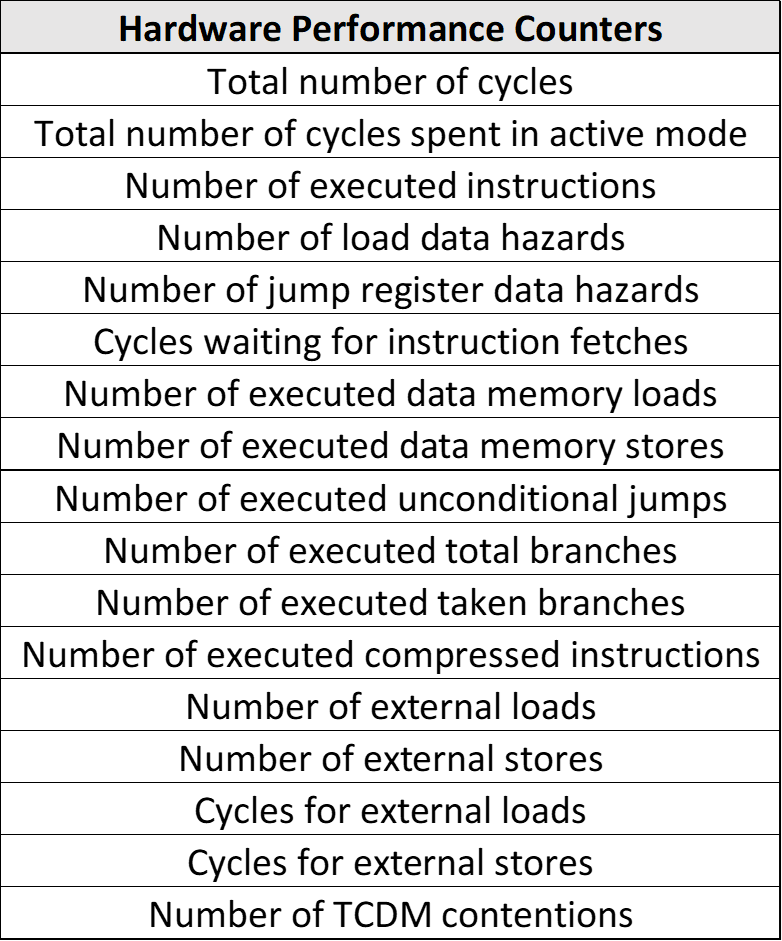}}
\caption{(a) GVSoC time management and clock domains synchronization. Example of request from FC to PULP CL L1 memory (b) GVSoC available performance counters.}
\label{fig:gvsoc_architecture/gvsoc_features}
\vspace{-0.25cm}
\end{figure*}
%%%%%%%% SYNCH %%%%%%%%
%
GVSoC falls into the event-driven simulator category with a hardware-oriented description based on PULP architecture. It can simulate a complete PULP platform modeling the micro-architecture and building common blocks such as cores, memory, cluster, peripherals, interconnect, TCDM.

GVSoC aims at allowing developers to test new architecture functionality (i.e., ISA extensions), to design micro-architecture and I/O drivers easily, and to explore design space for new chip features.
It also enables early-stage performance evaluation based on hardware counters and valuable debugging information to see how the components interact. It correctly simulates the timing across different clock domains such as PULP SoC, PULP CL, and peripherals, getting the precise performance.
It also supports different debugging tools, such as a test file of important architecture events and VCD traces, which can be dumped to help to debug the application and visualize application activity using profiling tools and optimize it.
\subsection{Structure}
\label{gvsoc_architecture/structure}
GVSoC includes three main components, as shown at the top of Fig.\ref{fig:gvsoc_architecture/gvsoc_structure}: C++ models, describing the behavior of the system components (e.g., core, memories, DMAs, peripherals, interconnect); configuration JSON files to configure the parameters of the architecture (e.g., bandwidth and latency of the interconnect), and a set of Python generators to instantiate all components of the specified target platform. This modular structure allows to compile the C++ models of the system components at the beginning and then to build a specific platform modifying the JSON files without recompiling the simulator and enabling fast DSE.
The JSON files describe the architecture to simulate and its modules, including the description of the FC, its main memory, the peripherals set, the PULP CL, how many Processing Elements (PEs) are in a PULP CL, TCDM size, cluster peripherals, and every element that summarize the particular chip to simulate.

All models are part of a library of components that can be assembled at runtime to build the system to be simulated.
The interaction between components is based on requests as shown in Fig.\ref{fig:gvsoc_architecture/gvsoc_req}. Requests contain every useful information to be shared with other modules such as memory-mapped location, payload, latency to accumulate and account for the current event, and the size of the reading or writing operation.
When a request comes from a component X to a component Y, the latter extracts the information it needs and responds with an appropriate returned value, depending on the nature of the request. If the request needs to be forwarded to another component, a specific method will be in charge of building the request and forwarding to one of its port. Every intermediate component could add latency to the initial request. The chain of requests will be concluded when a new event is generated or when the request reaches the receiver.
Main components of GVSoC are finely described in the following part.
\paragraph{CPU}
The \emph{CPU} component models the internal pipeline of a generic core with ports connected to the interfaces (instruction cache, logarithmic interconnect, and peripheral interconnect).
The pipeline includes an instruction fetch stage, a decode stage, and an execution stage. The instruction can specify a specific latency to modify the number of cycles required in the execution stage or a write-back latency related to data dependencies on the output register.
The main sub-component is the event-based ISS. The baseline RISC-V ISA and optional ISA extensions can be described using a configuration file that specifies the binary encoding and related metadata (e.g., instruction classes).
For each instruction, a callback function specifies its functional behavior and latency. The latency depends on the required execution cycles and the inferred pipeline stalls, which are modelled as events. Finally, the callback function also updates the internal core status (e.g., register values).
Platform designers can model a new instruction set by providing a tabular description of the instruction and the related callback functions.
\paragraph{Memory (TCDM / L2)}
The \emph{memory} component stores the data in an array and is accessible as word-data, half-word, or byte. Memories model the banks, which are configurable from configuration files, and every bank has a port to the \emph{interleaver}, a component that interfaces the banks with the interconnect. Internally, memories model the access requests from cores and DMAs as sequential events, setting a busy state with latency to serve the subsequent request. The requests of less or equal words are served in 0 latency if there are no contentions to the same memory bank. Simultaneously accesses or greater requests introduce latency.
\paragraph{DMA}
\emph{DMA} is a component that resides at the PULP CL domain and is specialized to autonomously move data from L1 memory to L2 and vice versa. It is directly connected to the memory via \emph{interleaver} and to the cores via dedicated ports. It is also connected to peripheral interconnect from PULP CL side and from SoC Domain to soc interconnect.
DMA has a memory region to be programmed, in which setting the size, the direction, and the memory locations of the transfer. Internally, it has a burst-based model and an associated latency, configurable from JSON files. The component splits the whole transaction in bursts of \emph{max\_burst\_size} and moves the data as an event that can be overlapped with other system events.
\paragraph{Interconnect}
There are two types of \emph{Interconnects} component in GVSoC. \emph{Interleavers} and \emph{Routers}, which model the interfaces of the interleaved memory and the routing of the generic non-word-level interleaved crossbars, respectively. The former models the routing from its ports (i.e., cores and DMA) to the TCDM memory banks, starting from the memory request address. The latter models the ports and have an associated latency to do the routing from one component to another. It statistically models the contentions, using the latency that will be incremented to the forwarded request like seen for Fig.\ref{fig:gvsoc_architecture/gvsoc_req}. For this reason, calibration is needed to better model the timing behaviour of the target architecture. Both the interconnect components are configurable from configuration files, specifying, for example, the number of ports, the latency, and the mapping.
\paragraph{I-Cache}
The \emph{I-Cache} model implements an n-ways set-associative cache adopting a least recently used (LRU) replacement policy. The configurable parameters are the number of ports, associativity, number of ways, and line size.
The cache is connected to an I-cache controller component, which connects the cache to the peripheral interconnect (to support data refill from the L2 memory) and to the cores (to support input, flush, and flush line requests).
\paragraph{Accelerators}
The accelerators are integrated both towards to the L1 interconnect and to the PULP CL's peripheral interconnect, exposing a memory-mapped programming interface.
The programming procedure involves setting the internal accelerator registers values and then triggering the beginning of a job.
Accelerators can support a queue of multiple jobs by employing a register shadowing mechanism.
When active, they fetch data directly from L1 with one or more dedicated ports on the interconnect to minimize contentions with the cores.
GVSoC already implements several accelerators and the standard interface to use as a template for new ones.
Adding a new accelerator requires specifying its address space, the name and the number of ports, and the events that it can raise.
\paragraph{I/O}
\emph{I/O} interfaces are grouped in the peripherals and micro-DMA domain. This component is connected to the L2, and the APB interconnect to be programmed by the core. GVSoC can be extended with a new peripheral modifying the proper configuration file, defining memory space, id and name of interfaces, and the chip selects. Then, the micro-DMA model instantiates the current peripheral, which are reachable from requests as the other components. Requests program the micro-DMA transactions, write/read internal register of specific peripheral, and kick-off the transfer between L2 and external device. Events are generated and executed at the peripheral speed, and synchronization is needed before fetch and store data from/to L2.
%
%%%%%%%% MOBILENET %%%%%%%%
\begin{figure*}[t!]
\centering
\includegraphics[scale=0.45]{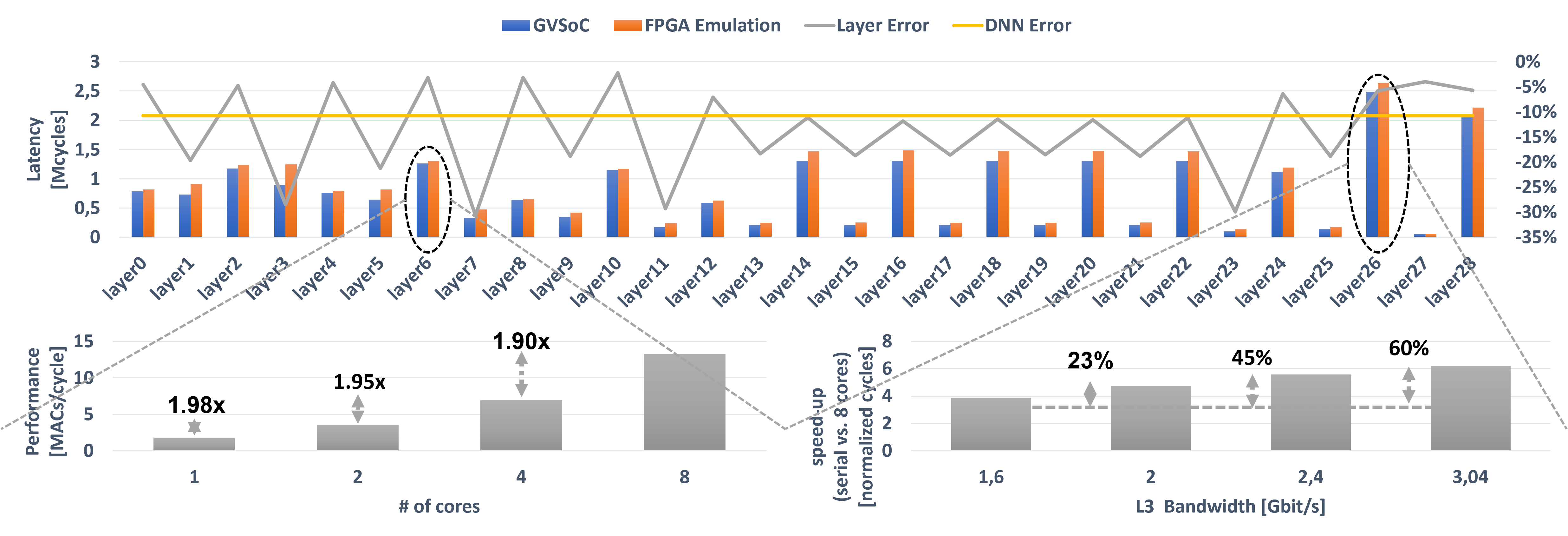}
\caption{MobileNetV1 execution cycles layer-by-layer and design space exploration on number of cores and L3 memory bandwidth for Layer6 and Layer26 respectively}
\label{fig:use_cases/mobilenet}
\vspace{-0.25cm}
\end{figure*}
%%%%%%%% MOBILENET %%%%%%%%
%
\subsection{Time Modeling}
\label{gvsoc_architecture/timing}
In addition to the functional top-accuracy, GVSoC has timing models for every relevant activity, such as instructions execution, DMA transfers, and memory accesses. It can emulate the actual system execution and provides a comprehensive set of statistics using hardware features or dedicated profiling tools.
In this context, a global \emph{time engine} manages the overall time (at the picosecond scale).
A \emph{clock engine} models a clock source as a forward monotone counter associated with a queue of related clock events, which are generic workloads associated with a specific clock cycle. Each event includes a data payload and a pointer to an associated callback function.
Clock engine defines a time window ($T_w$) in which the close enough events are included in a circular buffer as that in Fig.\ref{fig:gvsoc_architecture/time_modelling} for $Tw=8$. The execution of these events is done cycle-by-cycle, and simultaneous ones are executed sequentially using a non-ordered queue. The circular buffer can be fed with new events at any time if its execution cycle is inside $T_w$. If it is greater than $T_w$, the clock engine stores the event information in an ordered queue that is read every time a circular lap is completed. Fig.\ref{fig:gvsoc_architecture/time_modelling} shows the block diagram of the decision.
A different frequency can be set for each clock engine, enabling the integration with the global time engine. The mapping of clock events into the global time domain is performed by multiplying the clock period by the difference between the current clock counter and the clock time associated with the event.
As an additional level of modeling, all the components are related to a clock domain, which includes a clock engine and models the connections of the clock tree. A specific interface, called \emph{stub}, can be instantiated to synchronize the cycles if a request crosses two different clock domains. Fig.\ref{fig:gvsoc_architecture/gvsoc_clock_domain_cross} shows the stub interface between PULP SoC domain and PULP CL domain when a request from FC has to reach TCDM memory, and also the pseudo-code to convert the different cycle counts when a request is crossing a clock domain.
\subsection{Performance Assessment}
\label{gvsoc_architecture/performance}
To extract the execution and timing information, every core models a set of performance counters as the real hardware, measuring the events summarized in Fig.\ref{fig:gvsoc_architecture/counters}.
These counters, unlikely the actual hardware for a reason of area and power optimization, could be activated simultaneously, having one counter for each metric and saving a lot of simulation time.
Timing information is also stored in a complete set of system traces as shown at the bottom side of Fig.\ref{fig:gvsoc_architecture/gvsoc_structure} which provides information of what happens during the execution in every relevant module. Every trace could be either an event or if the event is quite critical, could there be more than one trace per event. The example of Fig.\ref{fig:gvsoc_architecture/gvsoc_structure} is referred to the same trivial example of C code beside, in which the core pipeline is shown in pseudo-code. The same system traces that could be used to extract timing information are beneficial to extract debugging information such as the content of the registers during the execution on FC or PULP CL cores and all the other events in the platform, such as DMA ids and pointers, L1, L2, and L3 memory accesses, cache event, event unit, and many others.
%
%%%%%%%%%%%%%%%%%%%%%%%%%%%%%%%% USE CASES %%%%%%%%%%%%%%%%%%%%%%%%%%%%%%%%%%%%%%
%
%%%%%%%% DSPAPPS %%%%%%%%
\begin{figure*}[t!]
\centering
\subfloat[\label{fig:use_cases/cores}]{\includegraphics[scale=0.37]{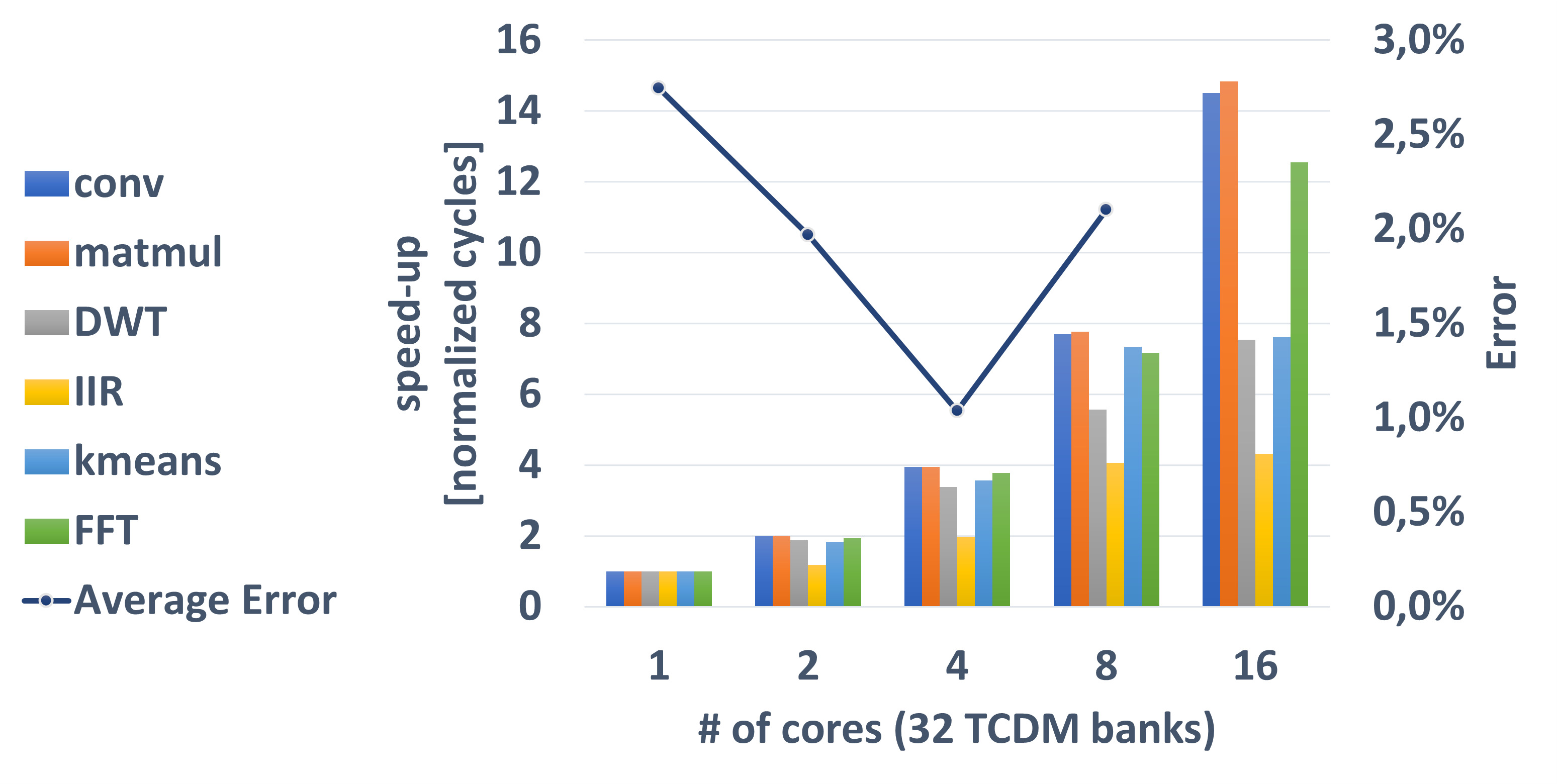}}
\hfill
\subfloat[\label{fig:use_cases/tcdm}]{\includegraphics[scale=0.37]{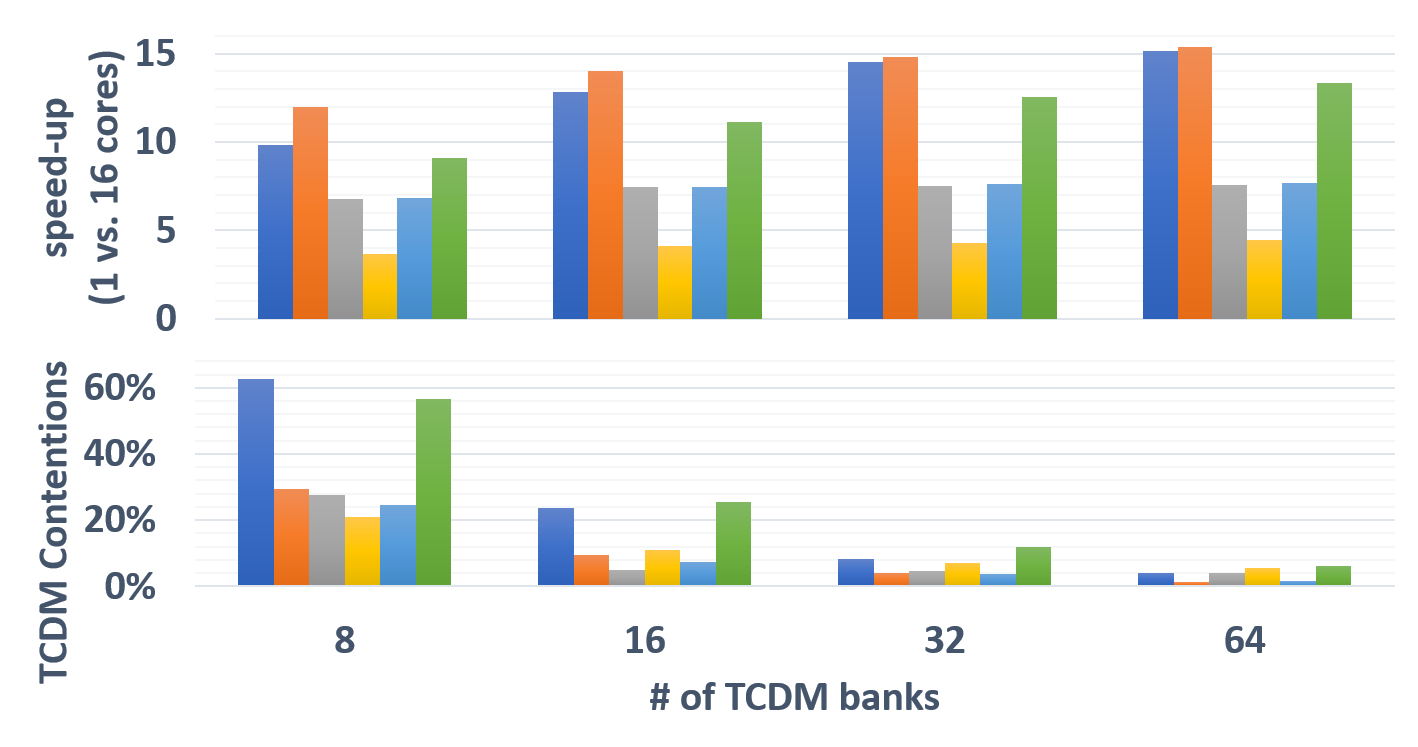}}
\caption{Common DSP kernel execution to show (a) the scalability with number of cores (b) how much TCDM banks impact the performance in relation with the TCDM contentions.}
\label{fig:use_cases/apps}
\vspace{-0.3cm}
\end{figure*}
%%%%%%%% DSPAPPS %%%%%%%%
%
\section{Use cases}
\label{sec:use_cases}
\begin{table}[t!]
\centering
\caption{Hardware performance counters in layer-by-layer MobileNetV1 execution }
\label{tab:use_cases/active_cycles}
\begin{tabular}{||c|c|c|c||}
    \hline
    \textbf{Layer}& \textbf{Total cycles}& \textbf{Active cycles}& \textbf{Difference}\\
    \hline\hline
    Layer6& 1263047& 1244504& 1\%\\
    \hline
    Layer26& 2481056& 1340781& 46\%\\
    \hline
  \end{tabular}
\vspace{-0.25cm}
\end{table}
%
%%%%%%%% ACCELERATOR %%%%%%%%
\begin{figure*}[t!]
\centering
\subfloat[\label{fig:use_cases/conv}]{\includegraphics[scale=0.37]{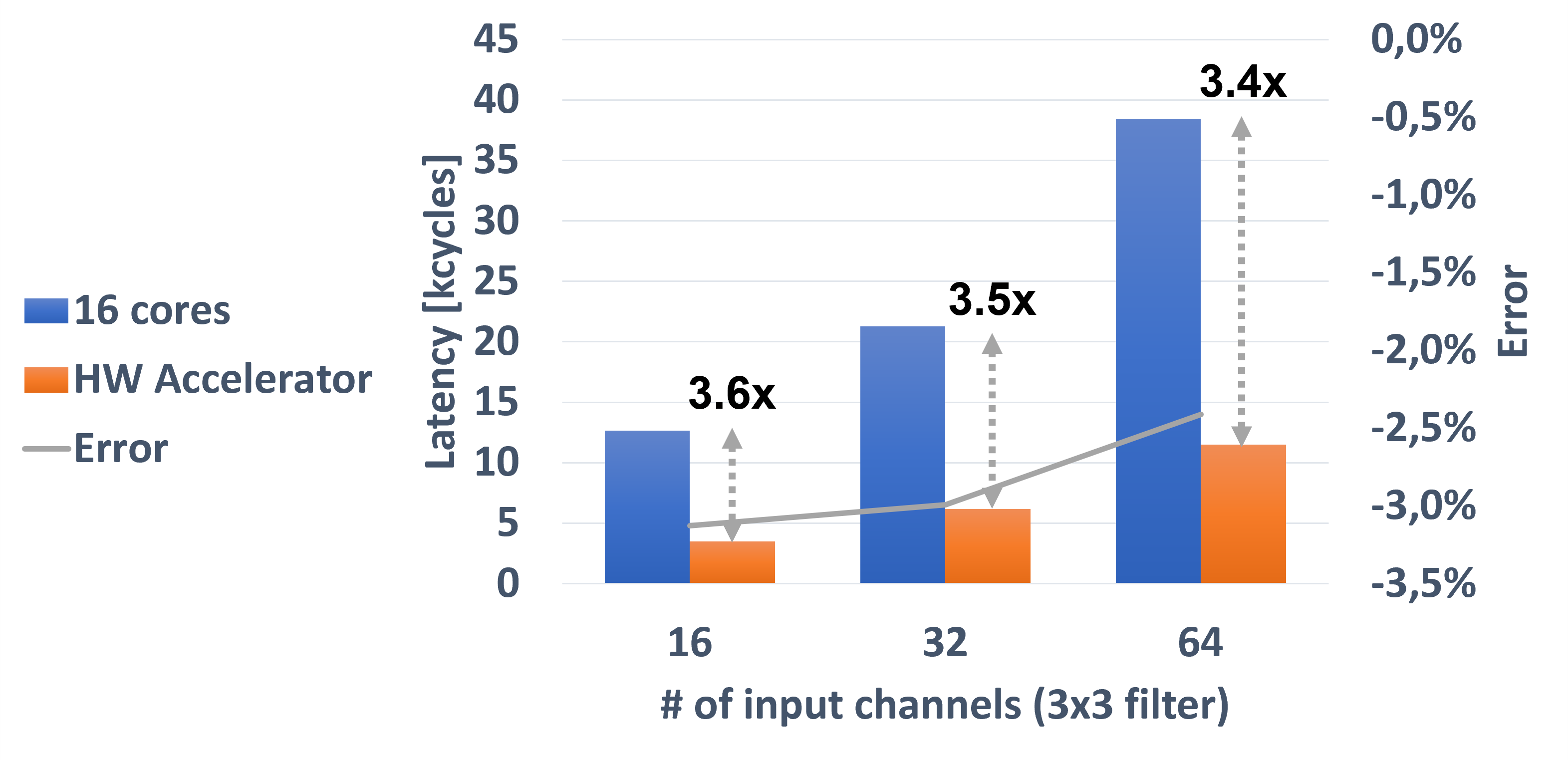}}
\hfill
\subfloat[\label{fig:use_cases/point}]{\includegraphics[scale=0.37]{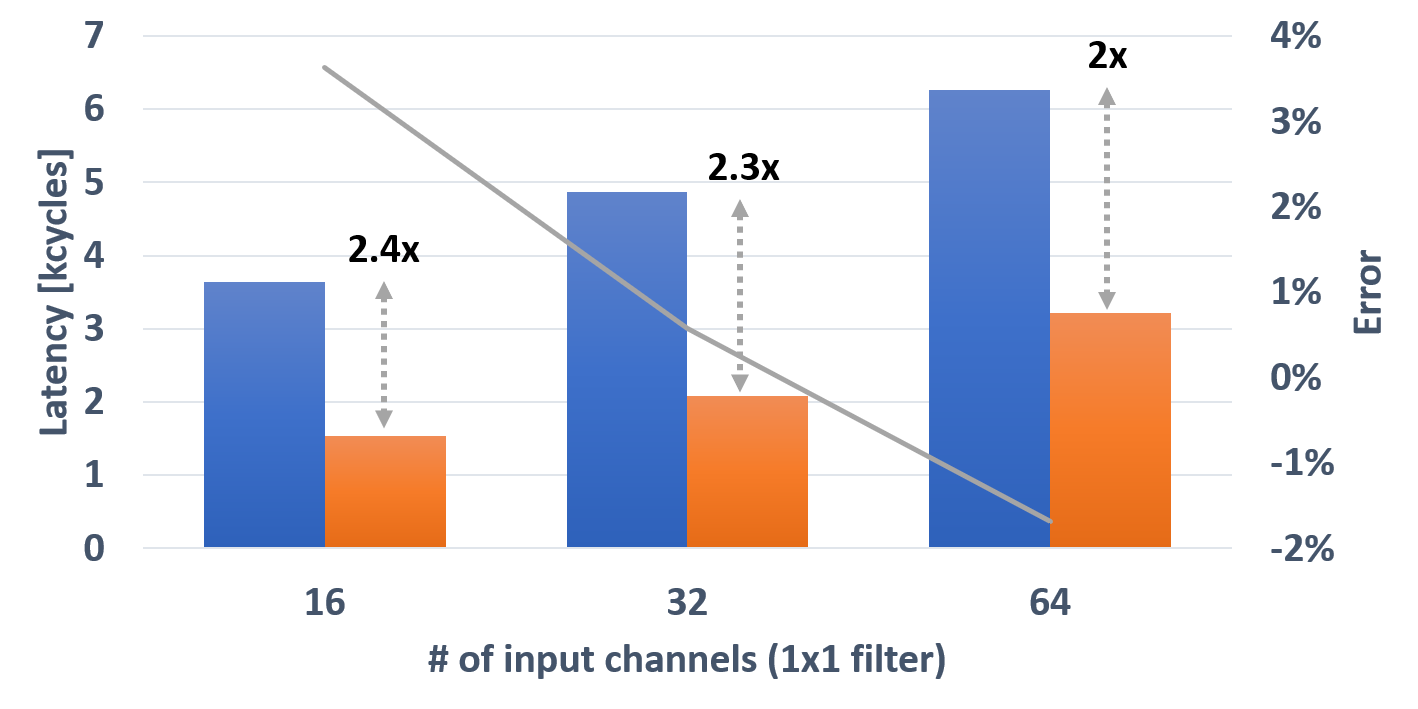}}
\caption{Comparison on convolution executions between 16 PULP CL cores and hardware accelerator varying $CH_{in}$ in: (a)  $K_{size}=3$ (b) $K_{size}=1$.}
\label{fig:use_cases/accelerator}
\vspace{-0.25cm}
\end{figure*}
%%%%%%%% ACCELERATOR %%%%%%%%
%
We take into account three use cases to demonstrate the suitability of GVSoC to deal with application optimizations, DSE of the key architectural parameters, and architecture structural design (e.g., including domain-specific accelerators). Firstly, we prototype the execution of a full MobileNetV1 \cite{howard2017mobilenets} showing the capabilities in terms of tracing, debug, and performance analysis enabled by execution on GVSoC; these features are not available on the real hardware, such as exploring the effect of the off-chip memory bandwidth on the computation performance, which is limited at 1.6 GBit/s. The second use case concerns the DSE of the TCDM banking factor, analyzing the execution of multiple common Digital Signal Processing (DSP) kernels.
Finally, we propose a comparison between an hardware accelerator for convolutions kernels and the execution on 16 PULP CL cores, with the aim to inspect the architecture limitations and bottlenecks.
\paragraph{MobileNetV1}
We employ the emulation of the target architecture on a Xilinx ZCU102, featuring an HyperRAM chip of 64 Mbytes of Flash and 8 Mbyte of DRAM, implementing the L3 level memory hierarchy.
The network is deployed by DORY \cite{burrello2021dory}, an open-source tool that allows automatic generation of optimized C-code for multiple state-of-the-art CNNs.
MobilenetV1 execution massively uses FC, L2 memory, micro-DMA, L3 HYPERRAM, PULP CL cores, DMA, and L1 TCDM.
To evaluate the accuracy with respect to the FPGA emulation, We have used the dedicated performance counter to get the total amount of cycles layer-by-layer, as shown at the top of Fig.\ref{fig:use_cases/mobilenet} with 8 PULP CL cores in Single Instruction Multiple Data (SIMD) fashion. In our experiment, we have observed an average error of around 10\%.
To provide more insights into the potential of the proposed simulator, we have performed DSE on a compute-bound and a communication-bound layer. In particular, we have focused on Layer6 and Layer26, two of the most expensive in terms of latency in the network.
As shown in Tab.\ref{tab:use_cases/active_cycles}, the difference between total cycles and active cycles for Layer6 is minimal. For this reason, we can conclude that performance is limited by computation and not by memory transfers. As shown at the bottom-left of Fig.\ref{fig:use_cases/mobilenet}, Layer6 potentially takes advantage of the parallelism of such computation, reaching a speed-up over the serial execution of 7.3$\times$.
Differently, Layer26 has a significant difference between the cycle counters; for this reason, we can deduce the communication limits the execution. This layer, in fact, stores weights and input/output activations in L3 memory, and thanks to the double-buffering approach, the code generated using DORY orchestrates the data transfers between different memory levels, up to PULP CL TCDM, overlapping as much as possible the data movements with core execution. However, if the PULP CL cores execute faster than DMA transfers, the cores enter sleep mode, waiting for the end of the data movement. This condition creates a gap between the two counters reported in Tab.\ref{tab:use_cases/active_cycles}.
As shown at the bottom-right of Fig.\ref{fig:use_cases/mobilenet}, increasing the bandwidth up to 3 GBit/s improves the speed-up metric up to 60\%.
\paragraph{DSP Applications}
The second use case explores how the number of cores and TCDM banks affects the execution on the PULP CL. As shown in Fig.\ref{fig:use_cases/cores}, bars represent the speed-up from 1 up to 16 cores, where the last is not designed for the FPGA emulation, with respect to 32 TCDM banks. The average error in these tests is below 3\%, and Fig.\ref{fig:use_cases/cores} shows how only convolution, matrix multiplication, and FFT current implementations can benefit from the highly parallel execution. The second experiment has been conducted considering the speed-up of having 16 PULP CL cores when the number of TCDM banks changes from 8 to 64. The speed-up of the best efficient execution layers improves by increasing the number of TCDM banks. This reduces the TCDM contentions, depicted in the bars at the bottom of Fig.\ref{fig:use_cases/tcdm}. Conversely, the other layers can not exploit the same benefit since they are not entirely parallelizable and TCDM contentions are structurally limited (i.e., the IIR filter in Fig.\ref{fig:use_cases/tcdm}, that reaches just 4$\times$ of speed-up over 16 cores).
\paragraph{Hardware Accelerators}
The third use case proposes an insight into a convolution hardware accelerator in PULP CL. We have analyzed the latency of different convolution operations to detect the architectural limitations of this type of accelerator with respect to an extremely parallel execution. Fig.\ref{fig:use_cases/accelerator} shows that the error of the GVSoC accelerator model is less than 4\%. The experiment has been conducted with 16 PULP CL cores with convolution dimensions of $CH_{in} \times 8 \times 8$, $K_{size} \times K_{size}$ and $32 \times 8 \times 8$ where we have varied $CH_{in}$ from 16, to 32 and then 64 and $K_{size}$ from 3 to 1. As shown in Fig.\ref{fig:use_cases/conv}, where the filter size is $3\times3$, the accelerator has a better response, reaching up to 3.6$\times$ less latency than the parallel execution. However, when the filter size decreases to $1\times1$, the accelerator performance drops, showing an architectural limitation.
This approach can quickly test new design parameters of the accelerator without implementing a complete hardware design.
%
%%%%%%%%%%%%%%%%%%%%%%%%%%%%%% STATE-OF-THE-ART %%%%%%%%%%%%%%%%%%%%%%%%%%%
%
\section{Comparison with the state-of-the-art}
\label{sec:state_of_the_art}
\begin{table*}[t!]
  \caption{Comparison between the state-of-the-art RISC-V simulators}
  \label{tab:soa/soa_table}
  \resizebox{\linewidth}{!}{%
  \begin{tabular}{||c|c|c|c||c|c|c|c||c|c||}
    \hline
    %\multicolumn{4}{||c||}{Type}&
    %\multicolumn{4}{|c||}{Support and Features}&
    %\multicolumn{2}{|c||}{Performance}\\
    %\hline
    \textbf{RISC-V} \textbf{Simulator}& \textbf{Target Field}& \textbf{Category}& \textbf{Open-Sourced}& \textbf{Core Extensions}& \textbf{Multicore}& \textbf{I/O Peripherals}& \textbf{Accelerators}& \textbf{Speed [MIPS]}& \textbf{Timing Accuracy}\\
    \hline\hline
    Spike\cite{monton2020riscv}& CPU& functional& yes& yes& yes& no& no& 170& n.d\\
    \hline
    RISCV-OVPSim\cite{ovpsim}& Full-platform& functional& yes& yes& yes& yes& no& 1000& n.d.\\
    \hline
    QuestaSim\cite{verilator_number}& all& cycle-accurate& no& yes& yes& yes& yes& 0.01& 100\%\\
    \hline
    Verilator\cite{verilator_number}& all& cycle-accurate& yes& yes& yes& yes& yes& 0.1& 100\%\\
    \hline
    gem5 (MinorCPU)\cite{butko2012gem5evaluation}& System& instruction-level& yes& no& yes& no& no& 0.2& 75\%\\
    \hline
    RISCV-TLM\cite{monton2020riscv}& Full-platform& event-driven& yes& no& no& yes& no& 8& unknown\\
    \hline
    RISCV-VP\cite{herdt2020riscvvp}& Full-platform& event-driven& yes& no& no& yes& no& 27& 95\%\\
    \hline
    & & & & & & & & & \\
    \hline
    \textbf{GVSoC}& \textbf{Full-platform}& \textbf{event-driven}& \textbf{yes}& \textbf{yes}& \textbf{yes}& \textbf{yes}& \textbf{yes}& \textbf{25}& \textbf{90\%}\\
    \hline
  \end{tabular}}
\vspace{-0.25cm}
\end{table*}
Just the most used RISC-V simulators have been considered and summarized in Tab.\ref{tab:soa/soa_table}, showing that GVSoC has a better trade-off between the simulation speed, accuracy, and the possibility of simulating complete and heterogeneous systems and platforms. Our experiment, to assess the simulation speed, is done on an Intel Core i7-8550U CPU at 1.88GHz and 16GB of memory.
As explained in Section \ref{sec:related_work}, functional simulators are typically complete and the fastest, but they do not model the timing and can not provide information about the execution. Cycle-accurate simulators are very slow, and GVSoC is up to 2500$\times$ faster than them, simulating all the system with high accuracy. Instruction-level simulators are not good enough in terms of accuracy and can not allow real design space exploration. The event-driven simulators like GVSoC simulate the entire platform but typically not model all key building blocks of a complete multicore SoC, such as heterogeneous accelerators, and I/O peripherals

\section{Conclusion}
\label{sec:conclusion}
In this work, we presented GVSoC, a configurable, fast and accurate simulator for IoT devices that enables design space exploration of real-world embedded devices. With its modular structure and configuration JSON files to describe the architecture and the components, Python-based generation scripts to build and instantiated the platform and the C++ IP models it can simulate several combinations of architectures and available resources simply changing parameters in the architecture description at run-time. For this reason, it can be used as a fast and robust simulator, reaching 25 MIPS and 10\% of mismatch between simulation and FPGA emulation timing results.

\section*{Acknoledgement}
This work was supported by the WiPLASH project (g.a. 863337) founded from the European Union’s Horizon 2020 research and innovation program.

\bibliography{bibliography}
\bibliographystyle{ieeetr}

\end{document}